\documentclass[epj]{svjour}

\usepackage[dvips]{graphicx}
\usepackage{bm,pstricks,longtable}
\usepackage{amsmath}
\usepackage{amssymb}
\usepackage{mathptmx}

\input{epsf}
\newcommand{\ket}[1]{| #1\rangle}                       %
\newcommand{\NP}{N_{\text{\tiny PAREC}}}		%
\newcommand{\US}{U_{\text{\tiny stat}}}		%

\begin{document}

\title{Quantum phase estimation algorithm in presence of static imperfections}
\author{Ignacio Garc\'ia-Mata\and Dima L. Shepelyansky}
\institute{\mbox{Laboratoire de Physique Th\'eorique,  
Universit\'e  de Toulouse III, CNRS, 31062 Toulouse, France}}
%\homepage[]{http://www.quantware.ups-tlse.fr/dima/}
%\mail{dima@irsamc.ups-tlse.fr}
\mail{http://www.quantware.ups-tlse.fr/dima/}
%\date{\today}
\date{November 12, 2007}

%%\begin{abstract}
\abstract{We study numerically the effects of static imperfections 
and residual couplings between qubits for 
the quantum phase estimation algorithm with two qubits.
We show that the success probability of the algorithm
is affected significantly more by static imperfections 
than by random noise errors in quantum gates.
An improvement of the algorithm accuracy can be reached
by application of the Pauli-random-error-correction method (PAREC).
}
%\end{abstract}

%\QICS{
%{11.80.+e}{Spectral evaluation}
%\and
%{13.10.+n}{Effects of noise and imperfections}
%\and
%{13.20.+e}{Quantum error correction}
%}

\maketitle
\section{Introduction}
Quantum computers \cite{chuang} are doing steady progress
increasing the number of qubits and accuracy 
of quantum gates. Among most advanced physical implementations
with possible scalable architecture are ion based
quantum computers (see e.g. \cite{blatt}
and Refs. therein) and solid state superconducting qubits
(see e.g. \cite{esteve,delft,yale}). 
In the present situation when only a few qubits are available
and the quantum gate accuracy is limited it is interesting to test
a performance of simple quantum algorithms 
operating at such realistic conditions.
One of such algorithms is the Iterative Quantum
Phase Estimation Algorithm (IQPEA) 
proposed recently by Dob\v{s}\'i\v{c}ek {\it et al.}~\cite{wendin}.
According to the results obtained there
the IQPEA works reliably even in presence of relatively strong 
noise in quantum gates. The algorithm \cite{wendin}
is based on the semiclassical Quantum Fourier Transform (QFT) 
\cite{griffiths} which 
uses one ancilla qubit, 
iterative measurements and a classical information feedback. 
The advantages of the algorithm are the following:
it uses only a single ancillary qubit and its  
theoretical accuracy of the eigenvalue found is limited only 
by the number of times the algorithm is applied. 
Due to that the IQPEA can be used as a benchmark algorithm for 
the maximal accuracy that could be 
obtained in a given experimental setup \cite{wendin}.

We describe briefly the IQPEA in the case of 
a two qubit circuit proposed in \cite{wendin}
as a minimal benchmarking circuit. The goal is to measure 
the eigenphase $\phi$ of some operator $U$ 
with precision set to $m$ significant bits. 
The standard Phase Estimation Algorithm (PEA) \cite{cleve98} 
requires $m$ ancillary qubits, to get the desired precision,
and the possibility to implement efficiently control-$U^{2^k}$ gates.
Using the semiclassical implementation of the QFT 
an alternative algorithm with only one ancillary qubit and
measurements can be designed \cite{kitaev}. One further step ahead 
is done  in \cite{wendin} where a feedback of the measurement result 
is used to correct the phase. In this way the IQPEA provides 
a way to compute the phase {\em theoretically\/}  with arbitrary 
precision. The proposed scheme can be used as a
benchmarking circuit which is tolerant  to a rather strong random noise 
in quantum gates. 

In \cite{wendin} only the case of random noise
errors in quantum gates and environment dephasing are considered.
At the same time it is known that a presence
of static imperfections and residual couplings
between qubits may lead to an emergence of quantum chaos
in a quantum computer hardware \cite{georgeot}.
Such static imperfections affect the accuracy of
quantum computation in a significantly stronger way
compared to random errors in quantum gates \cite{benenti2001,frahm}.
Thus it is interesting to test the effects of static imperfections
in IQPEA with a small number of qubits, e.g. two qubits.
Indeed, our studies presented in this paper
show that the static imperfections lead to a significant
drop of the computation accuracy and the algorithm success probability.
To correct these quantum errors induced by static imperfections in IQPEA
we apply the Pauli-random-error-correction (PAREC) method
proposed in \cite{parec2005} and tested in various quantum circuits
\cite{alber,viola}. Our results for IQPEA show that the PAREC allows to
improve significantly the accuracy of quantum computation.

The paper is organized as follows. First we briefly describe 
the IQPEA for a 2-qubit system (Section 2). We then compare the effects of
random phase errors in quantum gates and 
the effects of static imperfections (Section 3). 
Then we show how the PAREC method corrects the errors
induced by static imperfections (Section 4).
The summary of the results is given in Section 5.

\section{Brief description of IQPEA}
\begin{figure}[h!]
\begin{center}
\includegraphics[width=8cm]{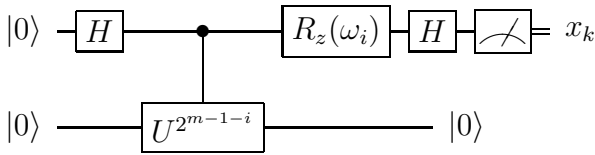}
\end{center}
\caption{Step $i$ (where $i=0,\,\ldots,\, m-1$) 
of the IQPEA of
\cite{wendin}. Here $R_Z$ is a rotation by an angle $\omega_i$ 
around $\hat{z}$ axis.
\label{fig:circ1}}
\end{figure}

The goal of IQPEA is to find the eigenphase of 
an operator $\hat{U}$. We consider the simplest case with 
one-qubit unitary diagonal operator in the computational basis
\begin{equation}
\hat{U}=\left(
\begin{array}{cc}
e^{-{\rm i}2\pi\phi}&0\\
0&e^{{\rm i}2\pi\phi}
\end{array}
\right),
\end{equation}
with $\phi\in[0,1]$. We want to find $\phi$ with up 
to $m$ bits of accuracy (or error smaller than $2^{-m}$).
The IQPEA \cite{wendin} 
procedure consists in applying $m$ times 
the circuit shown in Fig.~\ref{fig:circ1} to the 2-qubit state $\ket{00}$.
For simplicity we use the presentation
\begin{equation}
\phi=\sum_{i=1}^m \phi_i 2^{-i}\stackrel{\rm def}{=}
0.\phi_1\phi_2\ldots\phi_m 000\ldots 
\end{equation}
assuming that  the binary expansion of
$\phi$ is finite. For the first step ($i=0$) 
we take $\omega_i=0$ so the $Z$-rotation does not act.
After this first run  
\begin{equation}
\ket{\phi_0}=\frac{1}{2}\left[(1+e^{{\rm i}2 \pi\phi})\ket{00}+(1-e^{{\rm i}2 \pi\phi})\ket{10}\right]
\end{equation}
and the measurement  $\;\;$ of the ``left'' $\;\;$  qubit yields  $\;\;$
$P_0(\ket{0})=\cos^2(\pi(0.\phi_m))$ which 
is unity if $\phi_m=0$ and zero if $\phi_m=1$. 
Thus the least significant bit of $\phi$ is
obtained deterministically. The key element is that in  the following 
steps of the algorithm we use the classical information
obtained from the measurement to correct the phase by a Z-rotation. 
Before the last Hadamard gate the phase in the second step is
$2\pi(0.\phi_{m-1}\phi_m 00\ldots)$ and after performing a Z-rotation 
with $\omega_k=-2\pi(0.0\phi_{m-1})$ the probability becomes 
$P_1(\ket{0})=\cos^2(\pi(0.\phi_{m-1}))$. Consequently, the result of 
the first measurement is used as a feedback for the algorithm to
obtain the second least significant bit which is obtained deterministically. 
In theory, following this procedure each bit can be obtained.
%%%%%%%%%%%%%%%%%%%%%%%%%%%%%%%%%%%%%%%%%%%%%%%%%%%%%%%%%%%%%%%%%%%%%%%%%%%%
%%%%%%%%%%%%%%%%%%%%%%%%%%%%%%%%%%%%%%%%%%%%%%%%%%%%%%%%%%%%%%%%%%%%%%%%%%%%

In reality the phase is 
\begin{equation}
	\label{eq:phase}
\phi=\tilde{\phi}+\delta 2^{-m}
\end{equation}
where 
$\tilde{\phi}=0.\phi_1\phi_2\ldots \phi_m 000$ 
and $\delta\in [0,1)$ is the reminder. 
The probability of measuring $\phi_m$ correctly is thus
\begin{equation}
\left.
\begin{array}{l}
P_1(\ket{0})=\cos^2(\pi((0.\phi_{m})+\delta/2))\\
P_1(\ket{1})=\sin^2(\pi((0.\phi_{m})+\delta/2))
\end{array}
\right\}
=P_1=\cos^2(\pi\delta/2).
\end{equation}
The next step gives 
$P_2=\cos^2(\pi\delta/2^2)$ and eventually 
$P_k=\cos^2(\pi\delta/2^k)$ so that the total 
probability of measuring the phase correctly is given by
\begin{equation}
P_{\rm tot}(\delta)=\prod_{k=1}^{m}\cos^2(\pi\delta/2^{k})=
\frac{\sin^2(\pi\delta)}{2^{2m}\sin^2(\pi\delta/2^m)} \; .
\label{eq:prob_delta}
\end{equation}
The success probability in (\ref{eq:prob_delta}) 
is bounded in the limit $m\to \infty$ by $4/\pi^2$ \cite{cleve98}. 
In fact, the 
rounding error permits us to neglect 
the least significant bit and consider as probability of success the sum
$P_{\rm tot}(\delta)+P_{\rm tot}(1-\delta)=8/\pi^2$ 
when $m\to \infty$ \cite{wendin}. This lower bound could be raised 
by repeated measurement of the first few
bits and majority vote \cite{cleve98,wendin}. 
%[\texttt{In the paper they say that thus one needs $O[\log^2(1/\epsilon)]$ 
%extra measurements to get the correct
%result with error probability $\epsilon<1-8/\pi^2$ (independent of $m$).}]

\section{IQPEA and static imperfections} 
We consider two kinds of circuit imperfections: random phase errors 
in rotations and static imperfections due to residual
couplings between qubits.
To model random quantum phase errors we assume that the rotation  on angle
$\theta$  
\begin{equation}
R_{\sigma^{(\nu)}}(\theta)=\exp[-i \sigma^{(\nu)}\theta/2]
\end{equation}
(with $ \sigma^{(\nu)}$ a Pauli operator)
is replaced by rotation on angle $\theta(1+\Delta)$ with $\Delta$ 
randomly  and uniformly distributed in the interval
\begin{equation}
\Delta\in [-\frac{\epsilon_1}{2},\frac{\epsilon_1}{2}] \; .
\end{equation}
In other words the original rotation Hamiltonian has now an additional term
\begin{equation}
	\label{eq:rnd}
\delta H_{\text{\tiny rnd}}= \frac{\Delta\theta}{2}\,\sigma^{(\nu)}
\end{equation}
Each gate in Fig.~\ref{fig:circ1} is implemented with rotations 
having different random realizations of $\Delta$.
This is the case of random noise errors considered in \cite{wendin} 
where it was shown that the algorithm is
rather robust.
%%%%%%%%%%%%%%%%%%%%%%%%%%%%%%%%%%%%
\begin{figure}[h!]
\begin{center}
\includegraphics[width=8cm]{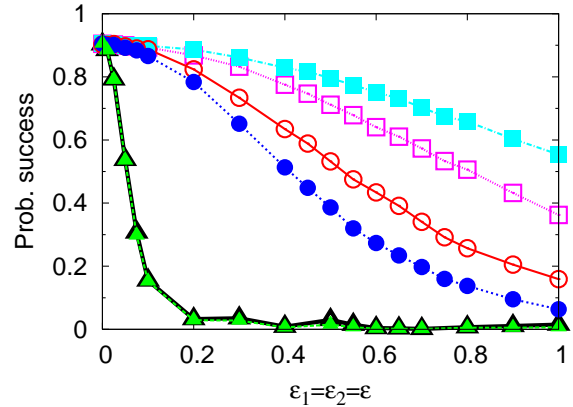}
\end{center}
\caption{(Color online) Success probability for 
the algorithm to determine the phase
with a precision of up to 10 bits, as a function of the parameter
$\epsilon=\epsilon_1=\epsilon_2$
characterizing the error strength. Symbols mark:
random phase errors in the Hadamard gates($\square$),
the $R_z$ gate ($\blacksquare$),  
the controlled-$U^{2^k}$ ($\circ$) and 
errors in all of the gates ($\bullet$). 
The case of static imperfections 
is shown by ($\triangle$)/($\blacktriangle$) in absence/presence of
random phase errors in the gates. 
Averaging is done over 2000 randomly 
chosen phases.\label{fig:two} }
\end{figure}
%%%%%%%%%%%%%%%%%%%%%%%%%%%%%%%%%%%%%%%%%%%%%%

On the other hand, we model the effects of residual static couplings 
between qubits by an imperfection Hamiltonian 
of the form used in \cite{georgeot,frahm}:
\begin{equation}
\label{deltaH}
\delta H_{\text{\tiny stat}}(x)=
\delta_1\sigma_1^{(z)}+
\delta_2\sigma_2^{(z)}+2 J\sigma_1^{(x)}\sigma_{2}^{(x)}
\end{equation}
where $\sigma_i^{(\nu)}$ are the Pauli operators acting on the $i$th qubit 
and $\delta_i,\,J$ are random coefficients uniformly distributed according to
\begin{equation}
\label{Jdelta}
%\delta_j=J_j\stackrel{{\rm def}}{=}\epsilon_j\in[\sqrt{3}
%\epsilon,\sqrt{3} \epsilon]\ .
\delta_i,\ J\in[-a\sqrt{3}
\epsilon_2,a\sqrt{3} \epsilon_2]\ ,
\end{equation}
(with $a$ constant). 
We suppose that between each gate in the algorithm there is a finite time
$\Delta t$ which remains fixed during the algorithm and 
that $\delta H$ acts via the unitary propagator
\begin{equation}
\US=e^{i \delta H} 
\end{equation}
where the time $\Delta t$ has been absorbed into the constants 
$\delta$ and $J$ in (\ref{deltaH}). The time $\Delta t$ can be considered as an effective
gate duration, a similar scheme is used in \cite{frahm}.

In order to compare the effects of both types of errors 
we compute $\langle{\rm tr}[\delta H^2]\rangle\propto \epsilon^2$. 
The value $a \approx 0.37$ is determined so that
\begin{equation}
\langle{\rm tr}[\delta H_{\text{\tiny rnd}}^2]\rangle\approx 
\langle {\rm tr}[\delta H_{\text{\tiny stat}}^2]\rangle\ \ {\rm if}\ \ \epsilon_1=\epsilon_2 \; ,
\end{equation}
the approximation is done taking $\theta =\pi$ in in Eq.~(\ref{eq:rnd}).
%%{\blue [THIS IS MAYBE A LITTLE BIT ARBITRARY\ldots]}.

The static type of imperfections is especially
important since generally the errors produced in this case are accumulated
coherently that leads to a 
quadratic term in the decay of fidelity \cite{frahm} thus 
limiting considerably the maximum time over which 
an accurate quantum computation can be performed.

%We take into account that the coupling between qubits can lead to change in the second qubit state, and if
%%before $P(\ket{0})\equiv P(\ket{00})$, now we have $P(\ket{0})\equiv P(\ket{00})+P(\ket{01})$.
In Fig.~\ref{fig:two} we show the success probability 
of measuring correctly the phase $\phi$, for
a chosen accuracy of $2^{-10}$, as a function of 
the parameters $\epsilon_1=\epsilon_2=\epsilon$. We
averaged over $2000$ uniformly random phases $\phi\in[0,1]$.  
For the cases where only random phase errors act we see 
that the algorithm is rather robust. Indeed, the decay
of the success probability is relatively slow
 even when all the gates involved have errors. 
On a contrary, a dramatic
drop of the accuracy of computation is seen 
when we include the effects of static imperfections. 
The decay is much faster than
for random phase errors even for comparable values 
of ${\rm tr}[\delta H^2]$ corresponding to a typical experimental situation.

%The small revival obseved in both static imperfection curves 
%is possibly due to some ammount of commutativity between the algorithm gates 
%gates and $\delta H_{\rm stat}$ [KLAUS].

%%%%%%%%%%%%%%%%%%%%%%%%%%%%%%%%%%%%%%%%%%%%%%%%%%%%%%
\begin{figure}[h!]
\begin{center}
\includegraphics[width=8.5cm]{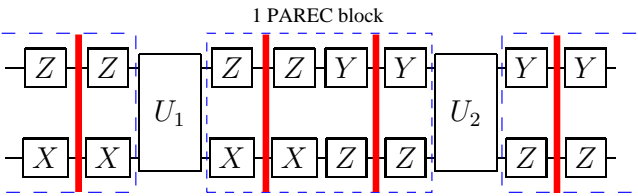} %%{pea.circ_2.eps}
\caption{(Color online) Schematic representation of the PAREC method. 
The vertical thick (red) lines indicate the
place where the static coupling propagator $\US$ is applied. 
In order to preserve
${\rm tr}[\delta H^2]$ we take $\epsilon/2$ for each propagator. 
The dashed lines enclose a PAREC block.
Repetition of PAREC between two gates means
applying repeatedly one PAREC block after another. \label{fig:parec}}
\end{center}
\end{figure}
%%%%%%%%%%%%%%%%%%%%%%%%%%%%%%%%%%%%%%%%%%%%%%%%%%%%%%

\section{Accuracy improvement using PAREC}
In this Section we address the issue of quantum error correction (QEC)
of errors induced by static imperfections during the
algorithm. The random errors in gates 
can be corrected up to a certain reasonable  
limit  by usual QEC schemes which however require
a significant increase of the number of qubits \cite{chuang}.
Here we study a different scheme to correct the effects produced 
by the static imperfections propagator $\US$.
One possible way to correct errors produced by residual static couplings 
was introduced in \cite{parec2005}. The idea is simple: contrary
to random errors the static imperfections
lead to a coherent accumulation of errors \cite{frahm}. 
If some randomness is introduced then the effect of the
residual couplings changes each time and does not accumulate coherently. 
The PAREC method \cite{parec2005} profits from the freedom of choice
of computational basis and uses this freedom by conveniently 
changing repeatedly and randomly the
computational basis along the computation. 
In order not to change the algorithm and the desired
result a special care must be taken to properly compensate for the changes made
renumbering the computational basis.
%%%%%%%%%%%%%%%%%%%%%%%%%%%%%%%%%%%%%%%%%%%%%%%%%%%%%%%%%%
\begin{figure}[h!]
\begin{center}
\includegraphics[width=9cm]{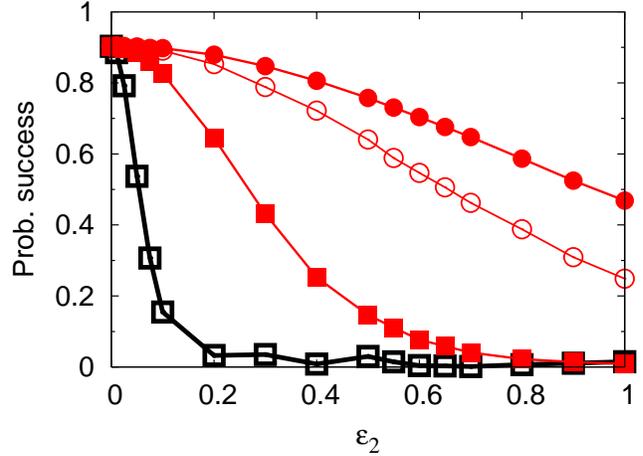}
\end{center}
\caption{(Color online) Success probability of IQPEA which
determines the phase
with a precision of up to 10 bits, in the presence of static imperfections, 
as a function of the imperfection strength $\epsilon_2$. 
The curves show the influence
of PAREC and how probability is enhanced with the increase of the number 
of times PAREC is applied: ($\square$) No PAREC; ($\blacksquare$) 1 time;
($\circ$)5 times; ($\bullet$) 10 times. No random imperfections 
are present ($\epsilon_1=0$). Averaging is done over 2000 randomly 
chosen phases.
\label{fig:four}}
\end{figure}
%%%%%%%%%%%%%%%%%%%%%%%%%%%%%%%%%%%%%%%%%%%%%%%%%%%%%%%%%%

The procedure is represented schematically in Fig.~\ref{fig:parec}. 
To change the computational
basis we first pick randomly from the set 
of Pauli matrices and identity matrix $\{ X_i,Y_i,Z_i,I_i\}$ 
(where $i=1,2$), 
and apply them to each qubit. We suppose that the time it takes 
to apply the Pauli operators is much
shorter than any other time scales.
We keep the information of this first choice, $X_1,Z_2$ in the Figure, 
and implement the suitably transformed gate 
$(Z_1\otimes X_2)U_1(Z_1\otimes X_2)$. After
that to come back to the original basis, 
the operator $(Z_1\otimes X_2)$ is applied again. The places
where $\US$ has acted are represented in Fig.~\ref{fig:parec} 
by thick vertical gray (red) lines.
This procedure is repeated before and after each gate, 
but of course the key is that a new
random sequence of Pauli operators is drawn, in the figure 
$(Y_1\otimes Z_2)$. So it is clear that 
between $U_1$ and $U_2$ the imperfection propagator 
$\US$ has acted on different bases.
We have called the complete sequence of Pauli operators 
that act between two gates one PAREC block. 

The effect of PAREC can be seen in Fig.~\ref{fig:four}. 
The black solid curve (with $\square$ symbols)
shows the success probability when static (but not random) imperfections act. 
The gray (red) line (with $\blacksquare$ symbols) 
show the result when one PAREC block is applied between each
gate of the algorithm that already gives a considerable gain.

\begin{figure}[h!]
\begin{center}
\includegraphics[width=9cm]{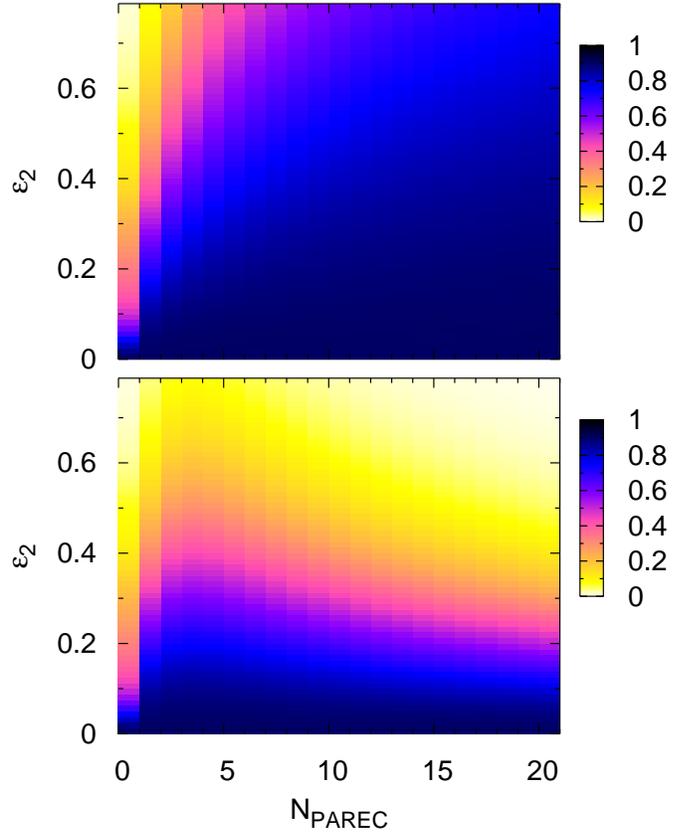} %{dens8_5.eps} %%{density2000.eps} %%
\caption{(Color online) Success probability 
(shown by color) of the IQPEA which determines the phase
with a precision of up to $2^{-10}$ as a function 
of the number of times $\NP$
PAREC is applied  and  of the static imperfections strength $\epsilon_2$. 
Top: the case where only static imperfections are considered 
($\epsilon_1=0$).
Bottom: both static and random imperfections are present 
(including in PAREC gates);
here random errors in gates are also present, their strength 
is taken as $\epsilon_1 =\epsilon_2/5 $. 
Averaging is done over 2000 randomly 
chosen phases.\label{fig:five}}
\end{center}
\end{figure}

 If instead of one PAREC block we introduce many 
of them ($\NP$) keeping $\Delta t$ fixed, and
supposing that the time to implement the Pauli gates is negligible, 
then the imperfection Hamiltonian
can be described with the help of the transformation
\begin{eqnarray}
\delta_i \rightarrow \frac{\delta_i}{\sqrt{2 \NP}} \; ; \; 
J \rightarrow \frac{J}{\sqrt{2 \NP}}.
\end{eqnarray} 
As a consequence, the coherent effect of static imperfections is suppressed. 
This can be seen in Fig.~\ref{fig:four}. The gray (red) curves show
the success probability as a function of $\epsilon_2$ 
for different values of $\NP$ (up to $\NP=10$). 
As $\NP$ grows the probability grows
accordingly. In the ideal limit of infinitely many 
PAREC blocks between gates (with a fixed gate-to-gate time) 
the success probability tends to constant maximum value for 
all $\epsilon$, a result which reminds
us of a Zeno-like effect \cite{misra77}. 
This is also illustrated in  Fig.~\ref{fig:five} (top), 
where the dark region in
the density plot of the success probability indicates 
the limiting value attained for large values of $\NP$.
The maximum value is the ideal value with perfect gates
which is only limited by the value of the reminder 
$\delta$ defined in (\ref{eq:phase}). Nevertheless, 
the limit can be attained only theoretically 
because the time between IQPEA gates cannot be
fixed if we add (ideally) infinitely many PAREC gates, 
no matter how fast we can implement them.

Up to now we have considered PAREC with perfect Pauli gates
while now we turn to a more realistic situation. 
With this aim  we also consider the possibility of
random phase errors of IQPEA gates to be also 
present in the PAREC Pauli gates.
Therefore we expect that in the presence of random imperfections, 
both in the IQPEA  and in PAREC, there will be
an optimal value of $\NP$ after which the presence of too 
many faulty Pauli gates yields PAREC useless. This is demonstrated
in Fig.~\ref{fig:five} (bottom).
For this a further consideration must be made. In Fig.~\ref{fig:two}, 
for illustration reasons only, we took $\epsilon_1=\epsilon_2$
such that the strength of both effects is comparable. 
However, we expect that in experiments the effect of random phase
errors can be reduced to a minimum, so that in fact 
we have to assume $\epsilon_2>\epsilon_1$. For the plot in 
Fig.~\ref{fig:five} we took $\epsilon_2=5\,\epsilon_1$. 
As a result the maximum of the success
probability occurs approximately at $\NP=5$ after which
the PAREC method looses its efficiency. 
The position of the peak as well as its height 
depends on the ratio $\epsilon_2/\epsilon_1$.
The obtained data show that the IQPEA with PAREC can 
operate reliably even in presence of relatively strong 
static imperfections.
%%%%%%%%%%%%%%%%%%%%%%%%%%%%%%%%%%%%%%%%%%%%%%%%%%%%%%%%%%%%%%%%%%%%%%%%%%
%%%%%%%%%%%%%%%%%%%%%%%%%%%%%%%%%%%%%%%%%%%%%%%%%%%%%%%%%%%%%%%%%%%%%%%%%%%

\section{Summary}
To summarize, 
we tested the effects of static imperfections in the IQPEA \cite{wendin}. 
Due to its simplicity
this algorithm can be used
as a benchmarking circuit for quantum computers with two qubits. 
We have shown that static imperfections produce 
a dramatic drop of success probability
even for algorithms involving a rather small number of gates.
In this context we  have tested the PAREC method \cite{parec2005} and shown that
it improves significantly the computation accuracy, 
even if the method is more suited to algorithms
with a larger gate sequence involved.  
We also present results with 
repetitions of the PAREC method 
that produces  a Zeno-like effect in
preservation of probability. Even though, the realistic 
scenario would suggest a small $\NP$ (may be even $\NP=1$), the
results obtained demonstrate a convincing improvement of the 
algorithm success probability induced by PAREC.
The extension of the IQPEA circuits to a larger 
number of qubits is straightforward, as well as the PAREC
implementation.

\section{Acknowledgments}
This work was supported in part by the EC IST-FET project $\;\;\;$
EuroSQIP. For numerical simulations we used the codes of 
Quantware Library \cite{qwlib}.


\begin{thebibliography}{10}
\bibitem{chuang} M.~A.~Nielsen, and I.~L.~Chuang,
                 {\it Quantum computation and quantum information},
                 Cambridge Univ. Press, Cambridge  (2000).
\bibitem{blatt} H.~H\"affner, W.~H\"ansel, C.~F.~Roos, J.~Benhelm, 
       D.~Chek-al-kar, M.~hwalla, T.~Körber, U.~D.~Rapol, M.~Riebe, 
       P.~O.~Schmidt, C.~Becher, O.~G\"uhne, 
       W.~D\"ur and R.~Blatt, Nature {\bf 438}, 643 (2005).
\bibitem{esteve} D.~Vion, A.~Aassime, A.~Cottet, P.~Joyez, H.~Pothier, 
       C.~Urbina, D.~Esteve, and M.H.~Devoret,
       Science {\bf 296}, 285 (2002).
\bibitem{delft} J.H.~Plantenberg, P.C. de Groot, 
       C.J.P.M.~Harmans, and J.E.~Mooij, Nature {\bf 447}, 836 (2007). 
\bibitem{yale} J.~Majer, J.M.~Cow, J.M.~Gambetta, J.~Koch, B.R.~Johnson,
       J.A.~Schreier, L.~Frunzio, D.I.~Schuster,
       A.A.~Houck, A.~Wallraff, A.~Blais, M.H.~Devoret, S.M.~Girvin,
       and R.J.~Schoelkopf, Nature {\bf 449}, 443 (2007).
\bibitem{wendin} M.~Dob\v{s}\'i\v{c}ek, G.~Johansson,
       V.~Shumeiko, and G.~Wendin, Phys. Rev. A {\bf 76}, 030306(R) (2007) .
\bibitem{griffiths} R.~B.~Griffiths and C.-S.~Niu, 
       Phys. Rev. Lett. {\bf 76}, 3228 (1996). 
\bibitem{cleve98} R. Cleve, A. Ekert, C. Macchiavello and M. Mosca, 
       Proc. R. Soc. Lond. A {\bf 454}, 339 (1998).
\bibitem{kitaev} A. Yu. Kitaev, Electron. Coll. Comput. Complex. 3 (1996)
                (arXiv:quant-ph/9511026 (2005)).
\bibitem{georgeot} B.Georgeot and D.L.Shepelyansky, 
       Phys. Rev. E {\bf 62}, 3504 (2000); {\bf ibid.} {\bf 62}, 6366 (2000).
\bibitem{benenti2001} G.~Benenti, G.~Casati, S.~Montangero and 
        D.~L.~Shepelyansky, Phys. Rev. Lett. {\bf 87}, 227901 (2001).
\bibitem{frahm} K.~M.~Frahm, R.~Fleckinger and D.~L.~Shepelyansky, 
        Eur. Phys. J. D {\bf 29}, 139 (2004).
\bibitem{parec2005} O.~Kern, G.~Alber, and D.~L.~Shepelyansky, 
        Eur. Phys. J. D {\bf 32}, 153 (2005).		
\bibitem{alber} O.~Kern and G.~Alber, Phys. Rev. Lett.
        {\bf 95}, 250501 (2005).
\bibitem{viola} L.~Viola and L.F.~Santos, J. Mod. Optics,
        {\bf 53}, 2559 (2006).
\bibitem{misra77}  B. Misra and E. C. G. Sudarshan, J. Math. Phys. {\bf 18}, 756 (1977);
    Wayne M. Itano, D. J. Heinzen, J. J. Bollinger, and D. J. Wineland, 
    Phys. Rev. A {\bf 41}, 2295 (1990);
       P. Facchi and S. Pascazio, Phys. Rev. Lett. {\bf 89}, 080401 (2002).
\bibitem{qwlib} K.~M.~Frahm and D.~L.~Shepelyansky (Eds.),
        {\it Quantware Library: Quantum Numerical Recipes},
        http://www.quantware.ups-tlse.fr/QWLIB/ .
\end{thebibliography}
\end{document}